\documentclass[10pt,twocolumn,journal]{IEEEtran}
\usepackage{latexsym}
\usepackage{amsfonts}
\usepackage{amsbsy}
\usepackage{amsmath,amssymb}
\usepackage{times}
\usepackage{graphicx}
\usepackage[usenames]{color}
\usepackage[dvips]{pstcol}
\title{Adaptive Minimum BER Reduced-Rank Interference Suppression Algorithms
Based on Joint and Iterative Optimization of Parameters}

\author{Yunlong Cai and Rodrigo C. de Lamare
\thanks{Y. Cai is with Department of
Information Science and Electronic Engineering, Zhejiang University,
Hangzhou 310027, China (e-mail: ylcai@zju.edu.cn). R. C. de Lamare
is with the Communications Research Group, Department of
Electronics, University of York, YO10 5DD York, U.K. (e-mail:
rcdl500@ohm.york.ac.uk).
This work is supported in part by the NSF of China
under grant $61101103$.
} }

\begin{document}
\maketitle \thispagestyle{empty}

\begin{abstract}
In this letter, we propose a novel adaptive reduced-rank strategy
based on joint iterative optimization (JIO) of filters according to
the minimization of the bit error rate (BER) cost function. The
proposed  optimization technique adjusts the weights of a
subspace projection matrix and a reduced-rank filter jointly. We
develop stochastic gradient (SG) algorithms for their adaptive
implementation and introduce a novel automatic rank selection method
based on the BER criterion. Simulation results for direct-sequence
code-division-multiple-access (DS-CDMA) systems show that the
proposed adaptive algorithms significantly outperform the existing
schemes.
\\


\end{abstract}

\section{Introduction}
 Detecting a desired user in a DS-CDMA system requires
processing the received signal in order to mitigate different types
of interference. In the scenario of interest of this work, the
receive filters are large, the interference is strong and
time-varying, and the training data provided to the receiver is
limited. In this context, reduced-rank signal processing has
received significant attention in the past years, since it provides
faster convergence speed, better tracking performance and an
increased robustness against interference as compared to full-rank
schemes operating with a large number of parameters. A number of
reduced-rank techniques have been developed to design the subspace
projection matrix and the reduced-rank filter
\cite{eign}-\cite{jidf}.  Among the first schemes are the
eigendecomposition-based (EIG) algorithms \cite{eign}, \cite{eign2},
the multistage Wiener filter (MWF) investigated in \cite{MWF2} and
\cite{MWF}, and the auxiliary vector filtering (AVF) algorithm
considered in \cite{avf1}. EIG, MWF and AVF have faster convergence
speed  compared to the full rank adaptive algorithms  with a much
smaller filter size, but their computational complexity is high. A
strategy based on the joint and iterative optimization (JIO) of a
subspace projection matrix and a reduced-rank filter has been
reported in \cite{delamarespl07,delamaretvt10}, whereas algorithms
with switching mechanisms have been considered in \cite{jidf} for
DS-CDMA systems.

{  Most of the contributions to date are either based on the
minimization of the mean square error (MSE) and/or the minimum
variance criteria \cite{eign}-\cite{jidf}. However, since the
measurement of transmission reliability is the bit error rate (BER)
not the MSE, they are not the most appropriate  metric from a
performance viewpoint in digital communications, as reported in
\cite{mber2}-\cite{mber3}.} Design approaches that can minimize the
BER have been reported in \cite{mber2}-\cite{mber3} and are termed
adaptive minimum bit error rate (MBER) techniques. The work in
\cite{mber3} appears to be the first approach to combine a
reduced-rank algorithm with the BER criterion. However, the scheme
is a hybrid  between an EIG or an MWF approach, and a BER scheme in
which only the reduced-rank filter is adjusted in an MBER fashion.

In this letter, we propose adaptive reduced-rank techniques based on
a novel JIO strategy that minimizes the BER cost function. The
proposed strategy adjusts the weights of both the rank-reduction
matrix and the reduced-rank filter jointly in order to minimize the
BER. By using multiple cycles over the recursions, we develop
stochastic gradient (SG) algorithms for an adaptive implementation
and introduce a novel automatic rank selection method with the BER
as a metric. Simulation results for DS-CDMA systems show that the
proposed algorithms significantly outperform existing schemes.

\section{DS-CDMA System Model}

Let us consider the uplink of an uncoded synchronous binary
phase-shift keying (BPSK) DS-CDMA system with $K$ users, $N$ chips
per symbol and $L_{p}$ propagation paths. The delays are multiples
of the chip duration and the receiver is synchronized with the main
path.
%
The
$M$-dimensional received vector is given by
\begin{equation}
\begin{split}
\mathbf {r}(i) & =  \sum_{k=1}^{K}A_{k}b_{k}(i) \mathbf {
\widetilde{p}}_{k}(i)+\mbox{\boldmath$\eta$}_{k}(i)+ \mathbf {n}(i),
\end{split}
\end{equation}
where $M=N+L_{p}-1$, $b_{k}(i)$ $\in$ $\{\pm1\}$ is the $i$-th
symbol for user $k$, and the amplitude of user $k$ is $A_{k}$. The
$M \times 1$ vector $\mathbf { \widetilde{p}}_{k}(i)=\mathbf
{C}_{k}\mathbf {h}_{k}(i)$ is the effective signature sequence for
user $k$, the $M \times L_{p}$ convolution matrix $\mathbf {C}_{k}$
contains one-chip shifted versions of the spreading code of user
$k$:
\begin {displaymath}
\mathbf {C}_{k}= \left( \begin{array}{ccc} a_{k}(1) &  & \mathbf{0} \\
\vdots &\ddots & a_{k}(1)\\
a_{k}(N) & &\vdots \\
\mathbf{0} & \ddots & a_{k}(N)
\end{array} \right),
\end{displaymath}
where $a_{k}(m) \in \{{\pm1}/\sqrt{N}\}$, $m=1,\ldots, N$.  The
channel vector of user $k$ is $\mathbf {h}_{k}(i)=[h_{k,0}(i) \ldots
h_{k,L_{p}-1}(i)]^{T}$, $\mbox{\boldmath$\eta$}_{k}(i)$ is the
inter-symbol interference (ISI), $\mathbf {n}(i) = [n_{0}(i) \ldots
n_{M-1}(i)]^T$ is the complex Gaussian noise vector with zero mean
and $E[\mathbf {n}(i)\mathbf {n}^{H}(i)] = \sigma^2 \mathbf {I}$,
where $\sigma^2$ is the noise variance,  $(.)^T$ and $(.)^H$ denote
transpose and Hermitian transpose, respectively.

\section{Design of MBER Reduced-Rank Schemes}

In this section, we detail the design of reduced-rank schemes which
minimize the BER. In a reduced-rank algorithm, an $M \times D$
subspace projection matrix $\mathbf{S}_{D}$ is applied to the
received data to extract the most important information of the data
by performing dimensionality reduction, where $1 \leq D \leq M$. A
$D \times 1$ projected received vector is obtained as follows
\begin{equation}
\mathbf {\bar{r}}(i)=\mathbf {S}^{H}_{D}\mathbf {r}(i),
\end{equation}
where it is the input to a $D\times 1$ filter $\mathbf
{\bar{w}}_{k}=[\bar{w}_{1}, \bar{w}_{2}, \ldots, \bar{w}_{D}]^T$. The
filter output is given by $\bar{x}_{k}(i)=\mathbf {\bar{w}}^{H}_{k}\mathbf
{\bar{r}}(i)=\mathbf {\bar{w}}^{H}_{k} \mathbf {S}^{H}_{D}\mathbf
{r}(i)$.
%
The estimated symbol of user $k$ is given by $\hat{b}_{k}(i)=\textrm {sign}\{\Re[\mathbf{\bar{w}}^{H}_{k}\mathbf{\bar{r}}(i)]\}$,
where the operator $\Re[.]$ retains the real part of the
argument and $\textrm {sign} \{.\}$ is the signum function.
The probability of error for user $k$ is given by
\begin{equation}
\begin{split}
P_{e} &= P(\tilde{x}_{k}<0)=\int^{0}_{-\infty}f(\tilde{x}_{k}) d\tilde{x}_{k}\\&= Q \bigg( \frac{\textrm{sign} \{ b_{k}(i)\}\Re[\bar{x}_{k}(i)]}{\rho (\mathbf{\bar{w}}^{H}_{k}\mathbf{S}_{D}^{H}\mathbf{S}_{D}\mathbf{\bar{w}}_{k})^{\frac{1}{2}}} \bigg),\label{eq:proberror}
\end{split}
\end{equation}
where $\tilde{x}_{k}=\textrm{sign} \{
b_{k}(i)\}\Re[\bar{x}_{k}(i)]$ denotes a random variable, $f(\tilde{x}_{k})$ is the single
point kernel density estimate \cite{mber2} which is given by
\begin{equation}
\begin{split}
f(\tilde{x}_{k})=&\frac{1}{\rho \sqrt{2\pi \mathbf{\bar{w}}^{H}_{k}\mathbf{S}_{D}^{H}\mathbf{S}_{D}\mathbf{\bar{w}}_{k} }}
 \\& \times \exp \bigg(
\frac{-(\tilde{x}_{k}-\textrm{sign}  \{ b_{k}(i)\}\Re[\bar{x}_{k}(i)])^2}{2\mathbf{\bar{w}}^{H}_{k}\mathbf{S}_{D}^{H}\mathbf{S}_{D}\mathbf{\bar{w}}_{k}\rho^2}\bigg),
\end{split}
\end{equation}
where $\rho$ is the radius parameter of the kernel density estimate,
$Q(.)$ is the Gaussian error function. The parameters of $\mathbf
{S}_D$ and $\mathbf{ \bar{w}}_{k}$ are designed to minimize the
probability of error. By taking the gradient of (\ref{eq:proberror})
with respect to $\mathbf {\bar{w}}^{*}_{k}$ and after further
mathematical manipulations we obtain
\begin{equation}
\begin{split}
\frac{\partial P_{e}}{\partial \mathbf{\bar{w}}^{*}_{k}} &
=\frac{-\exp \bigg(\frac{-|\Re[\bar{x}_{k}(i)]|^2}{2\rho^2 \mathbf {\bar{w}}^{H}_{k}\mathbf {S}_{D}^{H}\mathbf {S}_{D}
\mathbf {\bar{w}}_{k}} \bigg)\textrm{sign} \{ b_{k}(i)\}}{2\sqrt{2\pi}\rho}
\\ &\quad\times \bigg( \frac{\mathbf {S}^{H}_{D}\mathbf {r}}{(\mathbf {\bar{w}}^{H}_{k}
\mathbf {S}_{D}^{H} \mathbf {S}_{D} \mathbf
{\bar{w}}_{k})^{\frac{1}{2}}}-\frac{\Re[\bar{x}_{k}(i)] \mathbf
{S}^{H}_{D} \mathbf {S}_{D} \mathbf {\bar{w}}_{k}}{( \mathbf
{\bar{w}}^{H}_{k} \mathbf {S}_{D}^{H}\mathbf {S}_{D} \mathbf
{\bar{w}}_{k})^{\frac{3}{2}}}\bigg)\label{eq:proberror2}.
\end{split}
\end{equation}
By taking the gradient of (\ref{eq:proberror}) with respect to
$\mathbf {S}^{*}_{D}$ and following the same approach we have
\begin{equation}
\begin{split}
\frac{\partial P_{e}}{\partial \mathbf {S}^{*}_{D}} &
=\frac{-\exp \bigg(\frac{-|\Re[\bar{x}_{k}(i)]|^2}{2\rho^2 \mathbf {\bar{w}}^{H}_{k} \mathbf {S}_{D}^{H}\mathbf {S}_{D} \mathbf {\bar{w}}_{k}} \bigg)\textrm {sign}\{ b_{k}(i)\}}{2\sqrt{2\pi}\rho}
\\&\quad \times \bigg(\frac{\mathbf {r}\mathbf{ \bar{w}}^{H}_{k}}{(\mathbf {\bar{w}}^{H}_{k}\mathbf {S}_{D}^{H}\mathbf {S}_{D}\mathbf{ \bar{w}}_{k})^{\frac{1}{2}}}-\frac{\mathbf {S}_{D}\mathbf {\bar{w}}_{k}
\mathbf {\bar{w}}^{H}_{k}\Re[\bar{x}_{k}(i)]}{(\mathbf
{\bar{w}}^{H}_{k}\mathbf {S}_{D}^{H}\mathbf {S}_{D} \mathbf{
\bar{w}}_{k})^{\frac{3}{2}}} \bigg)\label{eq:proberror3}.
\end{split}
\end{equation}

\section{Proposed MBER Adaptive Algorithms}

In this section, we firstly describe the proposed scheme and MBER adaptive SG
algorithms to adjust the weights of $\mathbf {S}_D(i)$ and $\mathbf
{\bar{w}}(i)$ based on the minimization of the BER criterion. Then, a
method for automatically selecting the rank of the algorithm using
the BER criterion is presented.

\begin{figure}[!hhh]
\centering \scalebox{0.47}{\includegraphics{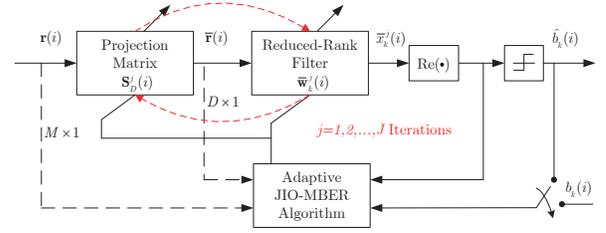}}
\vspace{-0.95em}\caption{Structure of the proposed reduced-rank
scheme } \label{fig:jiomberblock}
\end{figure}

\subsection{ Adaptive Estimation of Projection Matrix and Receiver}

The proposed scheme is depicted in Fig. \ref{fig:jiomberblock}, the
projection matrix $\mathbf {S}_{D}(i)$ and the reduced-rank filter
$\mathbf {\bar{w}}_{k}(i)$ are jointly optimized according to the
BER criterion. The algorithm has been devised to start its operation
in the training (TR) mode, and then to switch to the
decision-directed (DD) mode. The proposed SG algorithm is obtained
by substituting the gradient terms (\ref{eq:proberror2}) and
(\ref{eq:proberror3}) in the expressions $\mathbf{
\bar{w}}_{k}(i+1)=\mathbf{ \bar{w}}_{k}(i)-\mu_{w}\frac{\partial
P_{e}}{\partial \mathbf{ \bar{w}}^{*}_{k}}$ and $\mathbf{
S}_{D}(i+1)=\mathbf{ S}_{D}(i)-\mu_{S_{D}}\frac{\partial
P_{e}}{\partial \mathbf{S}^{*}_{D}}$ \cite{haykin} subject to the
constraint of $\mathbf{ \bar{w}}^{H}_{k}(i)\mathbf{ S}_{D}^{H}(i)
\mathbf{ S}_{D}(i)\mathbf{ \bar{w}}_{k}(i)=1$. 
%
%
Unlike prior JIO schemes, the proposed JIO-MBER algorithm employs
multiple cycles over the recursions for $\mathbf{ \bar{w}}_{k}$ and
$\mathbf{ S}_{D}$. At each time instant, the weights of the two
quantities are updated in an alternating way by using the following
equations
\begin{equation}
\begin{split}
\mathbf {\bar{w}}^{j+1}_{k}(i)&=\mathbf{
\bar{w}}^{j}_{k}(i)+\mu_{w}\frac{\exp
\bigg(\frac{-|\Re[\bar{x}^j_{k}(i)]|^2}{2\rho^2} \bigg)\textrm{sign}
\{ b_{k}(i)\}}{2\sqrt{2\pi}\rho}
\\&\quad\times \big( \mathbf{ S}^{j H}_{D}(i)\mathbf{ r}(i)-\Re[\bar{x}^j_{k}(i)]
\mathbf{ S}^{j H}_{D}(i)\mathbf{ S}^{j}_{ D}(i)\mathbf{
\bar{w}}^{j}_{k}(i)\big)\label{eq:proberror6}
\end{split}
\end{equation}
\begin{equation}
\begin{split}
\mathbf{ S}^{j+1}_{D}(i)&= \mathbf{ S}^{j}_{D}(i)+\mu_{S_{D}}
\frac{\exp \bigg(\frac{-|\Re[\bar{x}^{j}_{k}(i)]|^2}{2\rho^2}
\bigg)\textrm {sign}\{ b_{k}(i)\}}{2\sqrt{2\pi}\rho}
\\&\quad\times \big(\mathbf{ r}(i)\mathbf{ \bar{w}}^{jH}_{k}(i)-\mathbf{ S}^{j}_{D}(i)\mathbf{ \bar{w}}^{j}_{k}(i)
\mathbf{ \bar{w}}^{jH}_{k}(i)\Re[\bar{x}^{j}_{k}(i)]
\big)\label{eq:proberror7}
\end{split}
\end{equation}
where $\mu_{w}$ and $\mu_{S_{D}}$ are the step-size values, the
superscript $j$ denotes the $j$-th iteration at the time instant,
$j=1,\ldots, J$, and $J$ is the maximum number of iterations.
Expressions (\ref{eq:proberror6}) and (\ref{eq:proberror7}) need
initial values, $\mathbf {\bar{w}}^{1}_{k}(0)$ and $\mathbf
{S}^{1}_{D}(0)$, and we scale the reduced-rank filter by  $\mathbf{
\bar{w}}^{j}_{k}\leftarrow \frac{\mathbf{ \bar{w}}^{j}_{k}}{\sqrt{\mathbf{
\bar{w}}^{jH}_{k}\mathbf{ S}_{D}^{jH}\mathbf{ S}^{j}_{D}\mathbf{
\bar{w}}^{j}_{k}}}$ at each iteration.  The scaling has
an equivalent performance to using a constrained optimization with
Lagrange multipliers although it is computationally simpler. The
updated filters for the next time instant are given by  $\mathbf{
\bar{w}}^{1}_{k}(i+1)\leftarrow \mathbf{ \bar{w}}^{J}_{k}(i)$ and  $\mathbf{
S}^{1}_{D}(i+1)\leftarrow \mathbf{ S}^{J}_{D}(i)$. The proposed adaptive
JIO-MBER algorithm is summarized in table \ref{tab:table1}.

 \begin{table}[h]
\centering
 \caption{\normalsize Proposed adaptive JIO-MBER algorithms} {
\begin{tabular}{ll}
\hline
 $1$ & Initialize $\mathbf {\bar{w}}^{1}_{k}(0)$ and $\mathbf {S}^{1}_{D}(0)$.\\
 $2$ & Set step-size values $\mu_{w}$ and $\mu_{S_{D}}$ and the no. of iterations $J$.\\
 $3$ &   for each time instant $i$ do \\
 $4$ & ~~~ for $j$ from $1$ to $J$ do\\
 $5$ & ~~~~~~  Compute $\mathbf {\bar{w}}^{j+1}_{k}(i)$ and $\mathbf {S}^{j+1}_{D}(i)$ using (\ref{eq:proberror6}) and (\ref{eq:proberror7}). \\
 $6$ & ~~~~~~  Scale the $\mathbf {\bar{w}}^{j}_{k}$ using
  $\mathbf {\bar{w}}^{j}_{k} \leftarrow \frac{\mathbf {\bar{w}}^{j}_{k}}{\sqrt{\mathbf {\bar{w}}^{jH}_{k}\mathbf {S}^{jH}_{D}\mathbf {S}^{j}_{D}\mathbf {\bar{w}}^{j}_{k}}}$. \\
 $7$ & ~~~ end cycles \\
 $8$ &  After the $J$ cycles, obtain
 $\mathbf {\bar{w}}^{1}_{k}(i+1)\leftarrow \mathbf {\bar{w}}^{J}_{k}(i)$ \\&
and $\mathbf {S}^{1}_{D}(i+1)\leftarrow \mathbf {S}^{J}_{D}(i)$ for the next
time instant. \\
\hline
\label{tab:table1}
\end{tabular}
}
\end{table}
 Since the BER is employed as a design criterion, there is no
guarantee that the algorithms proposed will obtain the global
minimum solution \cite{mber2}. The proposed JIO-MBER technique and the other
existing BER-driven algorithms can only have their convergence
guaranteed to a local minimum.
In \cite{Niesen}, the work provides a general proof for the convergence of  alternating optimization algorithm.
By extending the work in \cite{Niesen}, we can obtain local minimum solutions for
the projection matrix and the reduced-rank filter of the proposed JIO-MBER algorithms.

\subsection{Computational Complexity of Algorithms and Delay Issues}
We describe the computational complexity of the proposed JIO-MBER
adaptive algorithm in DS-CDMA systems. In Table \ref{tab:table2},
 we
compute the number of additions and multiplications to compare the
complexity of the proposed JIO-MBER algorithm with the conventional  adaptive
reduced-rank algorithms, the adaptive least mean squares
(LMS) full-rank algorithm based on the MSE criterion \cite{haykin} and the SG
full-rank algorithm based on the BER criterion \cite{mber2}.
 Note that the
MWF-MBER algorithm corresponds to the use of the procedure in
\cite{MWF2} to construct $\mathbf{ S}_D[i]$ and
(\ref{eq:proberror6}) with  $J=1$ to compute $\mathbf{ \bar{w}}[i]$.
In particular, for a configuration with $N=31$, $D=6$ and $L_{p}=3$,
the number of multiplications for the MWF-MBER and the proposed
JIO-MBER algorithms  are $8377$ and $1262$, respectively. The number
of additions for them are $5920$ and $962$, respectively. Compared
to the MWF-MBER algorithm, the JIO-MBER algorithm reduces the
computational complexity significantly.

\begin{table}[h]
\centering%
\caption{\normalsize Computational complexity of Algorithms.} {
\begin{tabular}{ccc}
\hline \rule{0cm}{2.5ex}&  \multicolumn{2}{c}{Number of operations
per  symbol} \\ \cline{2-3}
Algorithm & {Multiplications} & {Additions} \\
\hline
\emph{\small \bf Full-Rank-LMS}  & {\small $2M+1$} & {\small $2M$}  \\
\emph{\small \bf Full-Rank-MBER}  &  {\small $4M+1$} & {\small $4M-1$}  \\
\emph{\small \bf MWF-LMS } \cite{MWF} &  {\small $DM^2-M^2$} &  {\small $DM^2-M^2$}  \\
                           &   {\small $+2DM+4D+1$}  &  {\small $+3D-2$} \\
\emph{\small \bf  EIG} \cite{eign2} &   {\small $O(M^3)$} &  {\small $O(M^3)$}  \\
\emph{\small \bf  JIO-LMS} \cite{delamarespl07} &   {\small $3DM+M$} &  {\small $2DM+M$}  \\
                          &   {\small $+3D+6$}  &   {\small $+4D-2$} \\
\emph{\small \bf MWF-MBER} \cite{mber3} &  {\small $(D+1)M^2$} &  {\small $(D-1)M^2$} \\
                       &   {\small $+(3D+1)M+3D$}  &  {\small $+(2D-1)M+2D$}          \\
                       &      {\small $+ML_{p}+10$}           &      {\small $+ML_{p}+1$}              \\
\emph{\small \bf JIO-MBER} & {\small $6MDJ+5DJ$}  & {\small $5MDJ+D J$} \\
                       &        {\small $+MJ+11J$}             &     {\small $-MJ-J$}            \\
\hline
\label{tab:table2}
\end{tabular}
}
\end{table}


Note that generally the delay associated with receive processing of filters is quite small as compared to the delay of channel decoding used in some systems such as maximum a posteriori (MAP), sum-product algorithm (SPA) and others  for turbo and LDPC codes.
We can find that the delay introduced by the receive filter is much lower than that of the channel decoder in \cite{delay1}.
We can see that the decoder has a much bigger impact on the delay and actually dominates the delay in the receiver \cite{delay2}.
For example, in the case of $64$ kbit/s UMTS transmission, the encoded broadcast service requires more than $5$ s system delay, and the uncoded systems such as conferencing and telephony require $150$ ms delay.

\subsection{Automatic Rank Selection}
The performance of reduced-rank algorithms depends on the rank $D$,
which motivates automatic rank selection schemes to choose the best
rank at each time instant \cite{MWF2,delamaretvt10,jidf}. Unlike
prior methods for rank selection, we develop a rank  selection
algorithm based on the probability of error, which is given by
\begin{equation}
P_{D}(i)=Q \bigg( \frac{\textrm{sign} \{
b_{k}(i)\}\Re[\bar{x}_{k}^{D}(i)]}{\rho }
\bigg)\label{eq:proberrorcostf}
\end{equation}
where the receiver is subject to $\mathbf{ \bar{w}}^{H}_{k} \mathbf{
S}_{D}^{H}\mathbf{ S}_{D}\mathbf{ \bar{w}}_{k}=1$. For each time
instant, we adapt a reduced-rank filter $\mathbf {
\tilde{\bar{w}}}_{k}(i)$ and a projection matrix $\mathbf {
\tilde{S}}_{D}(i)$ with the maximum allowed rank $D_{\rm max}$, which can be
expressed as $\mathbf{ \tilde{\bar{w}}}_{k}(i)=[\tilde {\bar{w}}_{1}(i), \ldots,
\tilde{\bar{w}}_{D_{\rm min}}(i), \ldots, \tilde{\bar{w}}_{D_{\rm
max}}(i)]^{T}$ and $\mathbf{ \tilde{S}}_{D}(i)=\left[ \begin{array}{ccccc} \tilde{s}_{1,1}(i)  &  \ldots & \tilde{s}_{1,D_{\rm min}}(i)
& \ldots &\tilde{s}_{1,D_{\rm max}}(i) \\
  \vdots &  \vdots & \vdots & \vdots &\vdots\\
\tilde{s}_{M,1}(i)  &  \ldots & \tilde{s}_{M,D_{\rm min}}(i) & \ldots &\tilde{s}_{M,D_{\rm max}}(i)\\
\end{array} \right]$
where $D_{\rm min}$ and $D_{\rm max}$ are the minimum and maximum
ranks allowed for the reduced-rank filter, respectively. For each
symbol, we test the value of rank $D$ within the range, namely,
$D_{\rm min}\leq D \leq D_{\rm max}$. For each tested rank, we
substitute the filter $\mathbf{
\tilde{\bar{w}}}^{'}_{k}(i)=[\tilde{\bar{w}}_{1}(i), \ldots,
\tilde{\bar{w}}_{D}(i)]^{T}$  and the matrix $\mathbf{ \tilde{S}}^{'}_{D}(i)=\left[ \begin{array}{ccc} \tilde{s}_{1,1}(i)  &  \ldots & \tilde{s}_{1,D}(i)  \\
  \vdots &  \vdots & \vdots \\
\tilde{s}_{M,1}(i)  &  \ldots & \tilde{s}_{M,D}(i) \\
\end{array} \right]$
into (\ref{eq:proberrorcostf}) to obtain the probability of error $P_{D}(i)$.
 The optimum rank can be selected as
\begin{equation}
D_{\rm opt}(i)=\arg \min_{D \in \{D_{min},\ldots, D_{max}\}}
P_{D}(i).
\end{equation}
 Note that after we obtain the projection matrix with rank
$D_{max}$, we generate the $M\times D$ projection matrix by grouping
the columns which are from the first column to the $D$-th column.
 The proposed auto rank selection mechanism only requires a
modest increase in the complexity of the JIO-MBER algorithm with a
fixed rank. The number of multiplications and additions for the
JIO-MBER algorithm with the auto rank mechanism are
$(6M+5)D_{max}+M+11$ and $(5M+1)D_{max}-M-1$, respectively. In
addition, a simple search over the values of $\bar{x}_{k}^{D}(i)$
and the selection of the terms corresponding to $D_{\rm opt}$ and
$P_{D_{\rm opt}}(i)$ are performed.

\section{Simulations}

 In this section, we evaluate the  performance of the
proposed JIO-MBER reduced-rank algorithms and compare them with
existing full-rank and reduced-rank algorithms.  Monte-carlo
simulations are conducted to verify the effectiveness of the
JIO-MBER adaptive reduced-rank SG algorithms. The DS-CDMA system
employs Gold sequences as the spreading codes, and the spreading
gain is $N=31$. The sequence of channel coefficients for each path
is $h_{k,f}(i)=p_{k,f}\alpha_{k,f}(i)(f=0,1,2)$. All channels have a
profile with three paths whose powers are $p_{k,0}=0$ dB,
$p_{k,1}=-7$ dB and $p_{k,2}=-10$ dB, respectively, where
$\alpha_{k,f}(i)$ is computed according to the Jakes model.
We optimized the parameters for the proposed and conventional
algorithms based on simulations. {  In this work, we focus on the
case with fixed step-size values and we note that, motivated by the
work in \cite{Kwong}, low-complexity variable step-size mechanisms
with the error probability as a metric can be further developed. }
%
The initial full rank and reduced-rank filters are all
zero vectors. The initial projection matrix is given by
$\mathbf{S}^1_{D}(0)=[\mathbf{I}_{D},\mathbf{0}_{D\times
(M-D)}]^{T}$.
The algorithms process $250$ symbols in TR and $1500$
symbols in DD. We set $\rho=2\sigma$.
%

\begin{figure}[!hhh]
\centering \scalebox{0.37}{\includegraphics{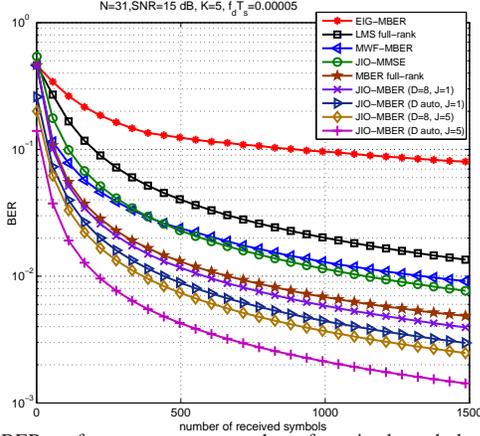}} \vspace{-2em}
\caption{ BER performance versus number of received symbols for the
JIO-MBER reduced-rank algorithms and the conventional schemes.
JIO-MBER (D=8, J=5): $\mu_w=\mu_{S_D}=0.005$. JIO-MBER (D auto,
J=5): $\mu_w=\mu_{S_D}=0.005$. JIO-MBER (D=8, J=1):
$\mu_w=\mu_{S_D}=0.105$. JIO-MBER (D auto, J=1):
$\mu_w=\mu_{S_D}=0.16$. EIG-MBER (D=8): $\mu_w=0.215$. LMS
full-rank: $\mu_w=0.105$. MBER full-rank: $\mu_w=0.05$. MWF-MBER
(D=8): $\mu_w=0.05$. ($D_{min}=3$, $D_{max}=20$)  }
\label{fig:simulation1}
\end{figure}

\begin{figure}[!hhh]
\centering \scalebox{0.37}{\includegraphics{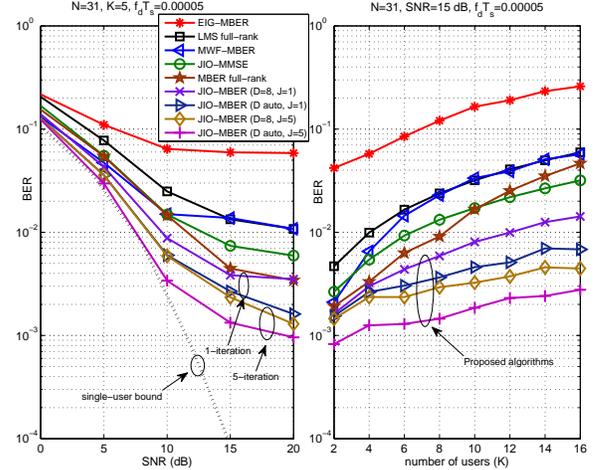}}
\vspace{-2em}\caption{  BER performance versus SNR and number of
users for the JIO-MBER reduced-rank algorithms and the conventional
schemes. $1500$ symbols are transmitted.  ($D_{min}=3$,
$D_{max}=20$) } \label{fig:simulation2}
\end{figure}

Fig. \ref{fig:simulation1} shows the bit error rate (BER)
performance of the desired user versus the number of received
symbols for the JIO-MBER adaptive SG algorithms and the conventional
schemes. We set the rank $D=8$, $K=5$, $SNR=15$dB and
$f_{d}T_{s}=5\times10^{-5}$. We can see that the JIO-MBER
reduced-rank algorithms converge much faster than the conventional
full rank and reduced-rank algorithms.
For the group of JIO-MBER
adaptive algorithms, the auto-rank selection algorithms outperform
the fixed rank algorithms.
%
Fig. \ref{fig:simulation2} illustrates the BER performance of the
desired user versus SNR and number of users $K$.
 In
particular, the  JIO-MBER algorithm using $D=8$ with
$J=1$ iteration can save up to over $6$dB and support up to six more
users in comparison with the MWF-MBER algorithm using $D=8$, at the
BER level of $2\times 10 ^{-2}$.


\section{Conclusions}

In this paper, we have proposed a novel adaptive MBER reduced-rank
scheme based on joint iterative optimization of filters for
 DS-CDMA systems. We have developed SG-based algorithms for
the adaptive estimation of the reduced-rank filter and the
projection matrix, and proposed an automatic rank selection scheme
using the BER as a criterion. The simulation results have shown that
the proposed JIO-MBER adaptive reduced-rank algorithms significantly
outperform the existing full-rank and reduced-rank algorithms at a
low cost.

\end{document}